\documentclass[journal]{IEEEtran}

\usepackage{cite}
\usepackage{amsmath,amssymb,amsfonts}
\usepackage{algorithmic}
\usepackage{textcomp}
\usepackage{multirow}
\usepackage{booktabs}
\usepackage{stfloats}
\usepackage{array}
\usepackage{wrapfig}
\usepackage{threeparttable}
\usepackage{color}
\usepackage{mathtools, cuted}
\usepackage{lipsum, color}
\usepackage{setspace}
\usepackage{stfloats}
\usepackage{graphicx}
\usepackage{footnote}
\usepackage{algorithm}
\usepackage{algorithmic}
\usepackage{gensymb}
\begin{document}

\title{Improving Intention Detection in Single-Trial Classification through Fusion of EEG and Eye-tracker Data}
%
%
%
\author{Xianliang~Ge$^*$,
        Yunxian~Pan,
        Sujie~Wang,
        Linze~Qian,
        Jingjia~Yuan,
        Jie~Xu,
        Nitish~Thakor,~\IEEEmembership{Fellow,~IEEE}, and~Yu~Sun$^*$,~\IEEEmembership{Senior Member,~IEEE}

\thanks{This work was supported by the National Natural Science Foundation of China (81801785, 31800931), the Hundred Talents Program of Zhejiang University, by the Zhejiang University Global Partnership Fund (100000-11320), by the Zhejiang Lab (2019KE0AD01), and by the Space Medical Experiment Project of China Manned Space Program (HYZHXM03001). (\emph{*, Corresponding author: X. Ge and Y. Sun})}
\thanks{X. Ge, Y. Pan and J. Xu are with the Center for Psychological Sciences, Zhejiang University, 310027, Hangzhou, Zhejiang, China, (X. Ge, email: 0918082@zju.edu.cn).}
\thanks{S. Wang, L. Qian, J. Yuan are with the Key Laboratory for Biomedical Engineering of Ministry of Education of China, Department of Biomedical Engineering, Zhejiang University, 310027, Hangzhou, Zhejiang, China.}
\thanks{N. Thakor is with the Department of Biomedical Engineering, Johns Hopkins University School of Medicine, United States, and also with the Department of Biomedical Engineering, National University of Singapore, 117456, Singapore.}
\thanks{Y. Sun is with the Key Laboratory for Biomedical Engineering of Ministry of Education of China, Department of Biomedical Engineering, Zhejiang University, and also with the Zhejiang Lab, 310027, Hangzhou, China, (email: yusun@zju.edu.cn).}
}
\maketitle

\begin{abstract}
Intention decoding is an indispensable procedure in hands-free human-computer interaction (HCI). Conventional eye-tracking system using single-model fixation duration possibly issues commands ignoring users’ real expectation. In the current study, an eye-brain hybrid brain-computer interface (BCI) interaction system was introduced for intention detection through fusion of multi-modal eye-track and ERP (a measurement derived from EEG) features. Eye-track and EEG data were recorded from 64 healthy participants as they performed a 40-min customized free search task of a fixed target icon among 25 icons. The corresponding fixation duration of eye-tracking and ERP were extracted. Five previously-validated LDA-based classifiers (including RLDA, SWLDA, BLDA, SKLDA, and STDA) and the widely-used CNN method were adopted to verify the efficacy of feature fusion from both offline and pseudo-online analysis, and optimal approach was evaluated through modulating the training set and system response duration. Our study demonstrated that the input of multi-modal eye-track and ERP features achieved superior performance of intention detection in the single trial classification of active search task. And compared with single-model ERP feature, this new strategy also induced congruent accuracy across different classifiers. Moreover, in comparison with other classification methods, we found that the SKLDA exhibited the superior performance when fusing feature in offline test (ACC=0.8783, AUC=0.9004) and online simulation with different sample amount and duration length. In sum, the current study revealed a novel and effective approach for intention classification using eye-brain hybrid BCI, and further supported the real-life application of hands-free HCI in a more precise and stable manner.
\end{abstract}


%
\IEEEpeerreviewmaketitle

\section{Introduction}
%
%
%
%
\IEEEPARstart{H}{ands}-free human-computer interaction (HCI) draws growing attention considering its convenience in various environments. As an efficient input modality in HCI, eye-gaze system was proposed in 1980s~\cite{ware1986evaluation,hutchinson1989human}. By the use of eye-tracker, selection is typically achieved through gazing on the target item for a specific period. Traditional eye-tracking system is straightforward to implement, ensuring short calibration time~\cite{jacob1991use} and faster target acquisition~\cite{stampe1995selection}. However, unexpected command was possibly issued when using the standalone eye-gaze system, leading to the Midas-touch problem~\cite{sibert2000evaluation,jacob1990you}. System cannot distinguish the spontaneous fixation from intended selection when dwelling time indeed exceeds the threshold. Besides, it is also inapplicable to adapt the threshold of dwell time to various ecological scenarios, so that the processing time of the user to the stimulus may overrun occasionally. Thus, in order to overcome this problem, other forms of input signals were introduced to assist the eye-gaze HCI system in intention decoding, such as special eye saccade and other physical movements~\cite{vspakov2012enhanced}. But such additional motor activities will induce extra mental workload and distract the execution of main task, as well as being inconvenient to the disabled. Therefore, in order to enhance the fluency and robustness of the system, a natural and intuitive input and decoding approach is necessary. 

Brain-computer interface (BCI) establishes a direct communication channel from brain signals~\cite{gao2021interface}, which can be used to detect the ongoing cognition, such as mental fatigue~\cite{jap2009using}, emotion state~\cite{dimitrakopoulos2018functional} as well as intention~\cite{atkinson2016improving,qi2018speedy,si2020predicting}. This technology has been proposed to decode the cognitive information implicitly from users' mind, without additional interruption to the primary task~\cite{zander2011towards}. Among the signal acquisition techniques of BCI, electroencephalography (EEG) has been widely adopted due to its numerous advantages in temporal resolution, usability and cost. As one of the most popular characteristics in EEG analysis, event-related potential (ERP) is generated by the neuron sensitive to the specified stimulus or events and broadly used to capture cognitive or sensory process~\cite{hillyard1983electrophysiology}. Generally, ERP includes several time-locked components (i.e. N170, a negative waveform at around 170 ms post-stimulus; P300, a positive waveform at around 300 ms post-stimulus, etc.) that correspond to particular cognition states~\cite{rugg1995erp}, which can serve as natural biomarkers of the user’s intention of interaction. This special characteristic of ERP has been evaluated for device control~\cite{li2016event}, target and error detection~\cite{protzak2013passive,donchin2000mental}, and so on. As a potential substitute for the “click” operation in HCI system, accumulating evidences confirmed that, when an intention of item selection emerged, negative potentials could be detected during conscious dwell time in the central~\cite{li2016event} and parietal~\cite{potts1996frontal,wiersema2007developmental} electrodes. One of the most popular paradigm in the EEG-based HCI research is P300 speller, which utilizes the uncommon event (flash of the target character) to induce P300 wave and decode user's intention~\cite{donchin2000mental}. As for the free visual search task, Kaunitz et al found that when subjects detected the target among distractors, a robust sensory component of fixation event-related potentials emerged, and a single-trial analysis could differentiate the type of stimulus based on EEG signals~\cite{kaunitz2014looking}. Devillez et al also observed a P300 component for fixation of the target natural scene compared with the free viewing without any target~\cite{devillez2015eye}.

In sum, the constituents of ERP contain abundant information towards personal intention and accompany gaze-based control intuitively in free search task, which can serve as the feature for distinguishing the intended selection from involuntary fixation. As a result, a combination of ERP and the eye-gaze input system could be complementary and provides more robust interaction experience. Accumulating evidence indicate that this hybrid BCI can satisfy the need for speed and accuracy simultaneously, overcoming the Midas-Touch problem of eye-tracker and inter-person variability of BCI protocol. For example, Kalika et al fused the eye-gaze data into a P300 speller pipeline and reported an improved classification accuracy and declined flash number for character spelling~\cite{kalika2017fusion}. Choi et al utilized the gaze position to contract a 12 $\times$ 12 character matrix into a 3 $\times$ 3 one, and highlighted this smaller navigation area to enhance the decoding performance of P300 speller~\cite{choi2013enhanced}. However, to the best of our knowledge, most hybrid BCI researches about the free visual search task only focus on the decoding of EEG signal, or take the EEG and eye-tracking for separate control purposes. Few of them attempted to analyze these two modalities in parallel and fuse them as input for intention classification.

\begin{figure*}[htbp]{\label1}
\centering
\includegraphics[scale = 0.95]{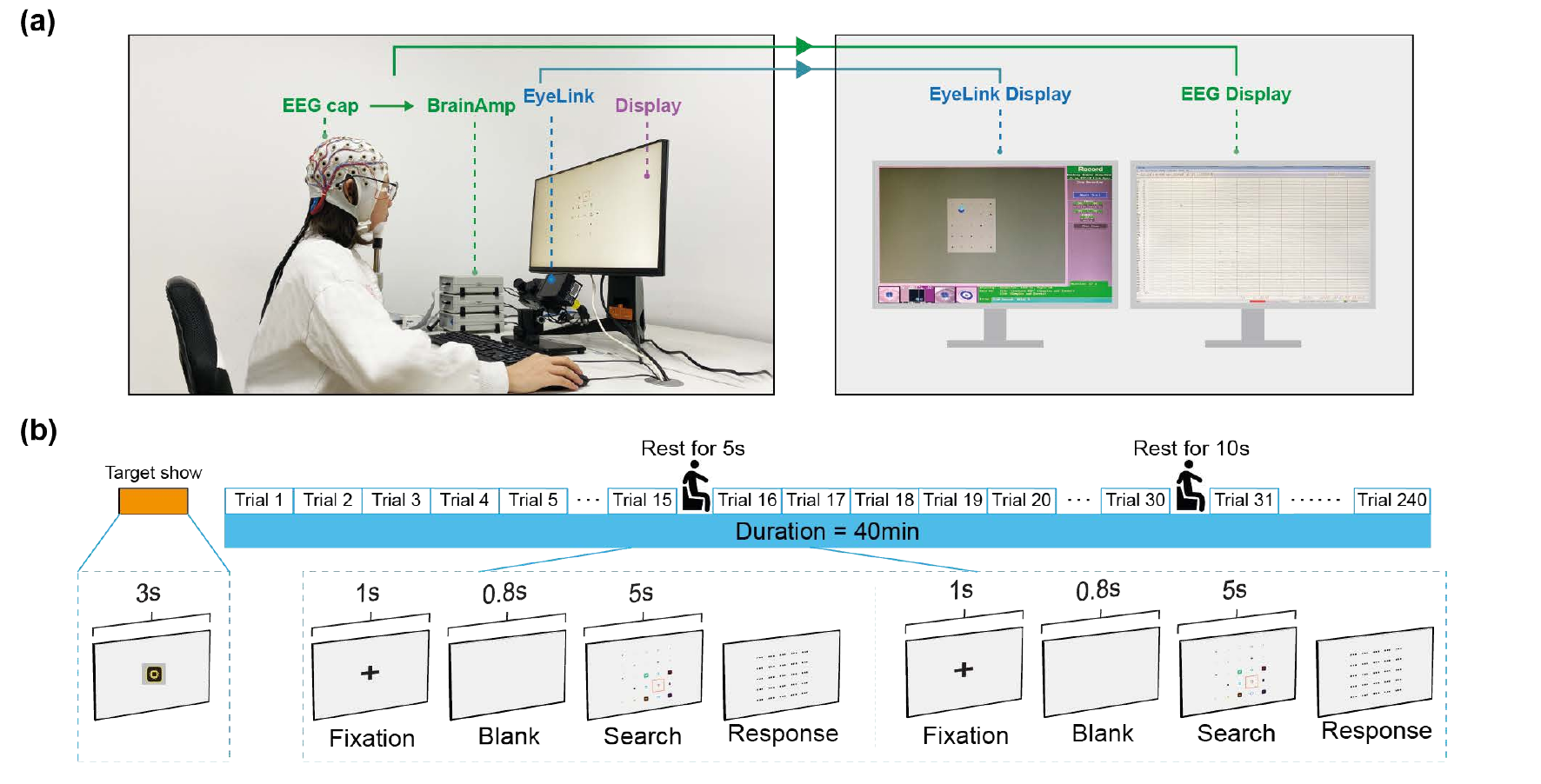}
\caption{A schematic diagram of the experimental protocol. (a) The setup of the experiment. EEG data was obtained from 64-channel BP system and eye-tracking data was collected by using EyeLink 1000 Plus system. (b) Each participant performed a 40-min target identification experiment, where the participant was asked to search the target icon as quick and accurate as possible and memorize and identify its location in the response period.}
\end{figure*} 

This research gap inspired our study to find out whether the alliance of input data from eye-tracker and BCI could facilitate the performance of intention detection in single-trial classification for active search task. Specifically, a self-designed HCI paradigm was proposed in which participants were required to search and identify a target icon among a total of 25 icons for each trial. We analyzed the fixation duration of each stimuli and corresponding ERP components, and these two inputs served as features for Target/Non-target classification. In the previous studies, LDA was extensively used for ERP detection owing to the satisfactory performance and simplicity~\cite{krusienski2008toward,martinez2016asynchronous}, while it also accompanied with the disadvantages of high noise sensitivity, poor inter-person generalization and the need of large training sample~\cite{lotte2018review}. Therefore, multiple improved LDA classifiers (including RLDA, SWLDA, BLDA, SKLDA, and STDA) with divergent edges were adopted for evaluating their performance in this task. And CNN, the most prevalent deep learning framework in the study of BCI~\cite{craik2019deep}, was also taken for comparison. Our study demonstrated that the fusion of concurrent ERP and fixation duration induced a superior performance over single feature in target intention decoding among all classifiers. Pseudo-online validation was further conducted to explore the proper amount of training set, response time of the system and optimal classification approach, so as to provide additional support for the practical application in various real-life HCI scenarios.

\section{Methods and materials}
\subsection{Subjects}
The study sample consisted of 70 university students (male / female = 35 / 35) from the Zhejiang University, China. All participants were aged between 17 and 29 years (mean age = 22.4 $\pm$ 2.3 years) and reported normal or corrected-to-normal vision. Participants with chronic illness, sleep disorder, childhood history of ADHD, and long-term medication history were excluded during pre-screened telephone interview. Prior to the experiment date, the included subjects were required to obtain a full night of sleep ($>$ 7 hours) for continuous 2 nights to minimise the effect of prior sleep restriction on neurobehavioral functions. Subjects consuming caffeine or alcohol, or undertaking strenuous exercise for 24 hour preceding the study were rescheduled. The study was approved by the Institutional Review Board of Zhejiang University and all participants signed informed consent prior to participation. 

\subsection{Experimental Protocol}
The experimental protocol was a typical target identification task that was customized using C programming language ({\color{blue}{Fig. 1}}). Specifically, participants were requested to search for a target icon among multiple non-target icons as quick and accurate as possible. A total of 25 icons were presented on a screen (1920 $\times$ 1080 pixels) with the background color was set at [R, G, B] = [192, 192, 192], which were arranged in 5 $\times$ 5. The experimental interface is shown in ({\color{blue}{Fig. 2}}). The size of each icon was set at 24 $\times$ 24 pixels, corresponding to a field of view (FOV) of 0.67$^{\circ}$ $\times$ 0.67$^{\circ}$. The horizontal/vertical distance between each pair of adjacent icons was set at 100 pixels. If an icon was highlighted, its size would be enlarged to 1.5 times of the original size (i.e., 36 $\times$ 36 pixels, FOV = 1$^{\circ}$ $\times$ 1$^{\circ}$. During the target searching, a certain icon which the participants gazed at and the surrounding eight icons would be highlighted ({\color{blue}{Fig. 2}}). 

\begin{figure}[htbp]{\label2}
\centering
\includegraphics[scale = 1]{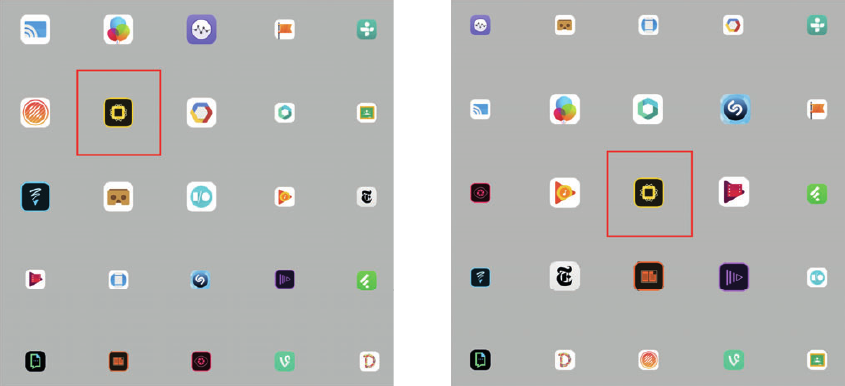}
\caption{Sample screenshots of visual target identification interface for two random trials. The icons surrounding the target icon as highlighted with a red box were enlarged to facilitate the identification.}
\end{figure} 

In the experiment, a predefined target icon (60 $\times$ 60 pixels, FOV = 1.67$^{\circ}$ $\times$ 1.67$^{\circ}$) was initially presented for 3 s. A black fixation was presented for 1 s, indicating the start of each experimental trial. Then, after a 0.8 s of a blank screen, the search interface was presented with a randomly assigned 5 $\times$ 5 icon pattern and a timer was started. The cursor was hidden at the moment. Participants were required to search the target icon within 5 s in the search interface. The mouse cursor appeared on the screen after finishing the 5 s searching period. Meanwhile, all the icons on the display were masked with a dotted line. The participant was requested to move the cursor to the location of the masked target icon and click to confirm the selection. The program would proceed to the next trial starting with a black fixation. A short period of break (i.e., 5 s) was introduced after completing 15 trials while a long period of break (i.e., 10 s) was introduced after completing 30 trials. For each participant, a total of 240 trials were administrated and the entire experiment lasted approximately 40 min.  

\subsection{Data Acquisition and Preprocessing}
The EyeLink 1000 Plus eye-tracking system (model: SR Research, Ottawa, Canada) was used to record the eye-tracking data. The sample rate was set at 1000 Hz. The participants were seated 60 cm from the monitor with a FOV of 13.86$^{\circ}$ $\times$13.86$^{\circ}$) using a chin support. Prior to the experiment, the eye-tracking system was calibrated for each participant. EEG data was recorded from a 64-channel EEG system (model: BrainAmp DC, Brain Products Gmbh, Gilching, Germany) according to the international 10-20 system. In addition, horizontal and vertical electrooculograms (EOG) were recorded on the lateral to the outer canthi (HEOG) as well as above and below (VEOG) the right eye. Electrode impedance was kept below 10 k$\Omega$ throughout the whole experiment. Anti-aliasing was achieved with a band-pass filter (0.5 $-$ 100 Hz) and additionally a 50 Hz notch filter was applied to avoid main interferences. Raw EEG and EOG signals were digitized at a sampling rate of 500 Hz using FCz as the reference. Two subjects were excluded due to data recording issues. 

In the analyzsis of eye-tracking data, we used the 100$\times$100 px area centered on each icon as an area of interest (AOI) of the icon. When a fixation falled in an AOI including non-target icon, this fixation was classified as the fixation of the non-target icon; while a fixaiton falled in an AOI including target icon, this fixation was classified as the fixation of the target icon. The fixation duration was calculated by summing up the duration of all fixations in the AOI.

A standard EEG preprocessing pipeline was adopted here, which included FIR band-pass filtering (1 $-$ 40 Hz), re-referencing to the average of all electrodes and ocular artifacts removal by removing the most correlated components to the EOG signals through independent component analysis (ICA)~\cite{jung2000removing}. All preprocessing steps were performed using customized codes and the EEGLAB toolbox~\cite{delorme2004eeglab} in Matlab 2017b (The MathWorks Inc, US). Greater details of the preprocessing steps could be found in our previous studies~\cite{dimitrakopoulos2018functional}.

The channel selection were based upon results in~\cite{krusienski2008toward}, which has been shown the optimal balance between classification performance and least number of electrodes. Here, Fz, Cz, Pz, Oz, P3, P4, PO7 and PO8 electrodes were included for the following classification. To extract features for target and non-target responses, the continuous EEG signals were segmented to 500 ms epochs with baseline correction by 100 ms interval before the icon presentation. Afterward each epoch was down-sampled to 32 Hz, that is 16 points for each channel and 128 points in total for all 8 channels.

\subsection{Classification Algorithms}
Several widely-used algorithms that were popular in the studies of ERP-BCI were adopted in the current work to assess the classification performance~\cite{xiao2019discriminative,blankertz2011single}, including regularized linear discriminant analysis (RLDA)~\cite{friedman1989regularized}, Stepwise linear discriminant analysis (SWLDA)~\cite{draper1998applied}, Bayesian linear discriminant analysis (BLDA)~\cite{lei2009empirical}, Shrinkage linear discriminant analysis (SKLDA)~\cite{vidaurre2009time}, Spatial-Temporal discriminant analysis (STDA)~\cite{zhang2013spatial}, and Convolutional neural network (CNN)~\cite{cecotti2010convolutional}. These algorithms were selected to cover the common categories of method for ERP-BCI, that is, concatenation of temporal points and spatial channels (RLDA, SWLDA, BLDA, SKLDA), adoption of spatial-temporal samples (STDA), and deep learning approach (CNN).  

\subsubsection{Regularized LDA} RLDA, a regularized version of LDA, is a popular technique for dimensionality reduction and feature extraction. It was originally introduced to solve the small sample size problem. The performance of RLDA technique depends upon the choice of the regularization parameter. In the current work, the regularization parameter was estimated using a deterministic approach according to~\cite{sharma2015deterministic}. This approach avoids the use of the heuristic cross-validation procedure for parameter estimation and improves the computational efficiency. Here the amount of regularization was set as $\lambda$ = 0.01. 
\subsubsection{Stepwise LDA} SWLDA, another regularized version of LDA, has been shown to be superior in the case of small sample size due to its implementation of combined forward and backward stepwise analysis to select suitable features in the discriminant model. Briefly, model estimation for SWLDA is conducted in a greedy manner by iteratively inserting and removing features from the model based upon statistical tests until the maximal number of active variable is reached or no additional features satisfy the entry/removal criteria. Here, the criteria was set as $p_{ins}$ = 0.1 and $p_{rem}$ = 0.15 as recommended in~\cite{krusienski2006comparison}. 
\subsubsection{Bayesian LDA} BLDA is a probabilistic method that based upon Bayesian regression and has been shown to outperform the original LDA method when only a small number of training sets was obtained or strong noise contamination in the data~\cite{hoffmann2008efficient}. According to~\cite{lei2009empirical}, the neurophysiological and experimental priors are employed explicitly by modeling the trial-level covariance and the weight vector covariance of LDA explicitly as linearly separable components with the relative contribution of each component is controlled by the hyperparameters that could be estimated via Restricted Maximum Likelihood.  
\subsubsection{Shrinkage LDA} Through adjusting the extreme eigenvalues of the covariance matrix towards the average eigenvalue, SKLDA improves the traditional LAD when using insufficient training samples. For high-dimensional data with only a few data points given (i.e., EEG data), the estimation for a covariance matrix may become imprecise, which may lead to a systematic error: large eigenvalues of the original covariance matrix are estimated too large, and small eigenvalues are estimated too small. Of note, shrinkage is a common remedy for compensating the systematic bias of the estimated covariance matrices and shrinkage parameter for high-dimensional feature spaces. In the current work, the shrinkage parameter was set at 0.1 according to~\cite{schafer2005shrinkage}. For details of SKLDA and its interpretation could be found in~\cite{blankertz2011single}
\subsubsection{Spatial-Temporal Discriminant Analysis} STDA is a multiway extension of the LDA that tries to maximize the discriminant information between target and nontarget classes through finding two projection matrices from spatial and temporal dimensions collaboratively. Unlike the abovementioned different versions of LDA method where data were concatenated as input, through incorporating the spatial and temporal information, STDA reduces the feature dimensionality in the discriminant analysis and decreases the number of required training samples~\cite{zhang2013spatial}. 
\subsubsection{Convolutional Neural Network} CNN was initially used in computer vision and has gained substantial interest in BCI most recently for its superior performance. In this research, a five-layer CNN was developed for EEG pattern detection. The input of the network was a 2D space-time EEG signal with a size of 8 $\times$ 16. It was followed by two paired layers, with each pair comprised a convolutional layer with batch normalization and a max-pooling layers. In the first convolutional layer, we utilized 32 kernels with a size of 1 $\times$ 5 for time domain convolution. While the second was used for spatial domain
convolution, containing 32 kernels with a size of 8 $\times$ 1, which equaled the number of EEG electrodes. After each convolution  process, a ReLU function was employed for non-linearization. For max-pooling layers, they both utilized a pooling filter size of 1 $\times$ 2 to reduce computational complexity. After dropout process with a dropout rate of 0.3, the output of max-pooling layer was applied to two fully connected layer comprising 64 and 2 neurons respectively. In the decision step, the classification probability is determined by softmax function.

\subsection{Offline Classification}
In order to demonstrate that fusion of EEG and eye-tracker data would lead to superior performance in comparison with single EEG or eye-tracker modality, classification was initially performed using only EEG or eye-tracker data. Specifically, fixation duration corresponding to extracted epoch of one gaze was initially estimated and selected as input for LDA classifier as a benchmark. For EEG data, a 0 $-$ 500 ms epoch after a gaze was cut out and selected as input for the offline classification. Of note, one trial might have multiple target and non-target samples (corresponding to the search for target) with the number of non-target samples larger than that of target samples. A cross-validation approach was initially employed to assess the performance of classifiers under different number of training samples using a reformatted balance data. Specifically, the training set (Target:Non-target = 1:1) was designed using sample number from 30 to 420 with a step of 30, while the testing set was randomly selected from the remaining samples and maintained a Target:Non-target = 1:1 fashion with a maximum amount. Of note, the same training and testing samples were applied on all classification algorithms to allow for fairness comparison. This procedure was repeated for 10 times, and the average area under curve (AUC) of the receiver operating characteristic (ROC) curves was computed for the quantitative comparisons. Then, a separate 10-fold cross-validation approach was applied on the real data (on average, Target:Non-target $\approx$ 1:2.3) to demonstrated that the fusion of multi-modal features induced a superior performance over single feature. 

\subsection{Pseudo-online Validation}
\subsubsection{Online Classification} A pseudo-online analysis was performed to validate the feasibility and practicability of implementing the decoding algorithm based upon our analysis framework. As at least 240 trials were performed for each subject in the experiment, the samples within the first 80 trials were utilized as training set whereas the remaining data were considered as testing set for assessing the performance of online classification. To avoid the imbalance of the sample amount between two classes in the training set and for the convenience of result analysis in the testing set, the number of the target and non-target classes was set to equal respectively. Of note, fixation duration longer than 500 ms would be redefined as 500 ms to ensure same duration of EEG data.
\subsubsection{Epoch threshold} As it has been mentioned previously, gaze fixation duration was defined as the duration post a gaze for either target or non-target and EEG data between 0 $-$ 500 ms was used as threshold for data extraction and the following classification. In order to assess the influence of different threshold on the classification, we have also used 300 ms to 800 ms with a step of 50 ms as threshold for the epoch extraction. For instance, for a predefined threshold (e.g., 400 ms), gaze fixation duration above the threshold would be redefined as the threshold value and the EEG data between 0 $-$ 400 ms would be used as input. Besides, the number of training trials was also put into consideration as a factor contributing to online performance. In detail, the samples within the first 20 to 100 trials with a step of 10 trials were regarded as training set, while the remaining were used for testing. The ratio of Target and Non-target was rearranged to 1:1 in both training and testing set as well. 
\begin{figure}[htbp]{\label3}
\centering
\includegraphics[scale = 0.875]{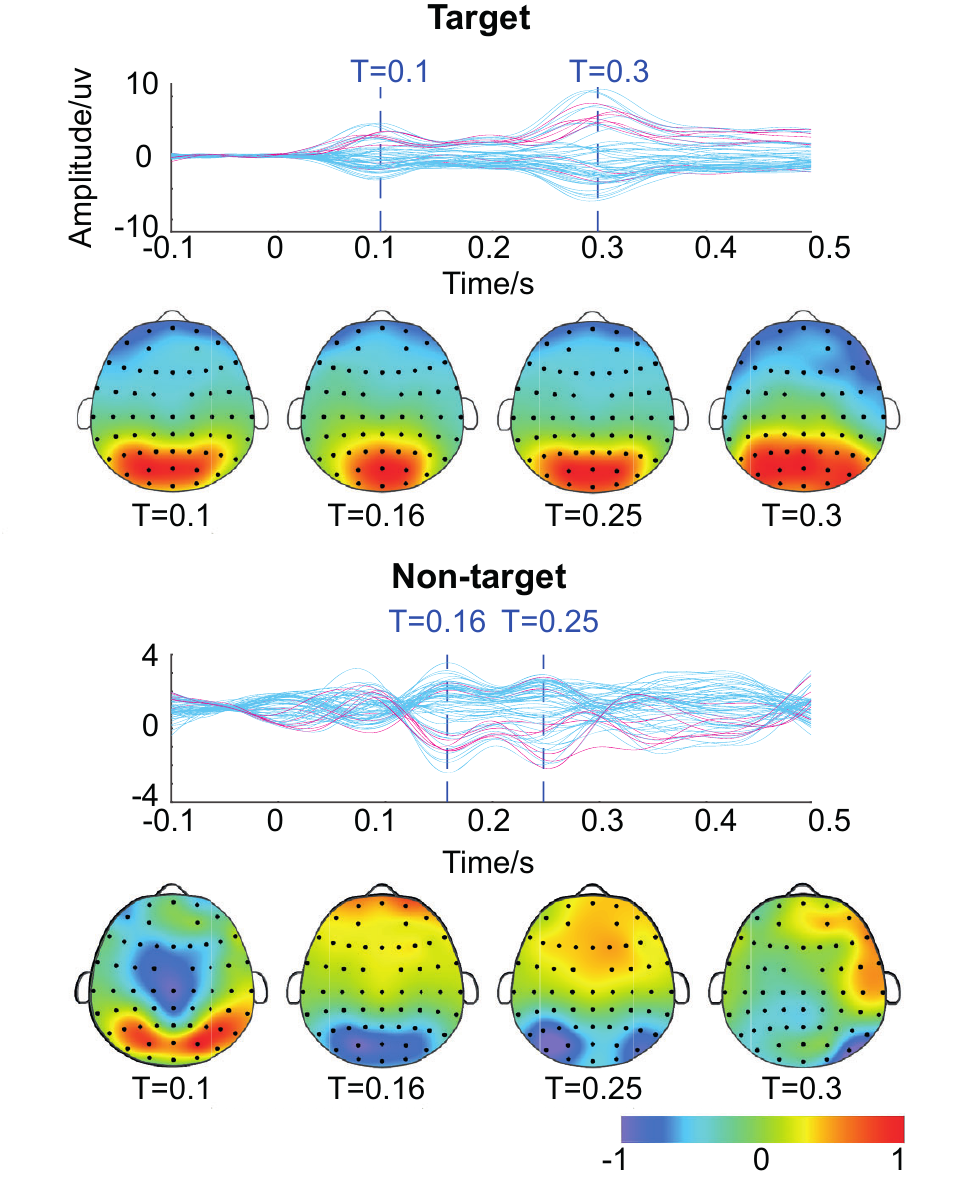}
\caption{Distributions of different ERP characteristics of Target and Non-target from a randomly-selected subject.}
\end{figure}
\section{Results}
\subsection{Behavioral Performance}
Data from four participants were excluded for signs of poor motivation on the task, likely due to boredom experienced during the target identification experiment. Threshold for signs of poor motivation on the task was set if the error rate of the participant was 1 S.D. lower than the group average. Our final dataset thus consisted 64 participants (male / female = 29 / 35) and the following classification was conducted on these participants. Overall, the remaining participants performed the experiment well, as indicated by the relatively high detection rate (mean $\pm$ S.D. = 98.12\% $\pm$ 1.33\%). We had performed additional statistical analysis to assess the gender effect and found insignificant difference between males and females (t$_{62}$ = 0.084, p = 0.933). 

\subsection{Characteristics of ERPs}
The characteristics of ERPs were first analyzed and compared between target and non-target stimulus. {\color{blue}{Fig. 3}} shows the temporal and spatial differences between two kinds of ERPs for a randomly-selected subject. Specifically, the discriminant ERP features between target and non-target were restricted to the occipital areas post-stimulus. Hence, these evident differences between target and non-target serve as salient underlying features for the following classification algorithms. Moreover, the observed posterior differences were in line with the findings in~\cite{krusienski2008toward} and justify the selection of the EEG channels (i.e., Fz, Cz, Pz, Oz, P3, P4, PO7 and PO8 in this work) for the classification algorithms. 

\begin{figure}[htbp]{\label4}
\centering
\includegraphics[scale = 1]{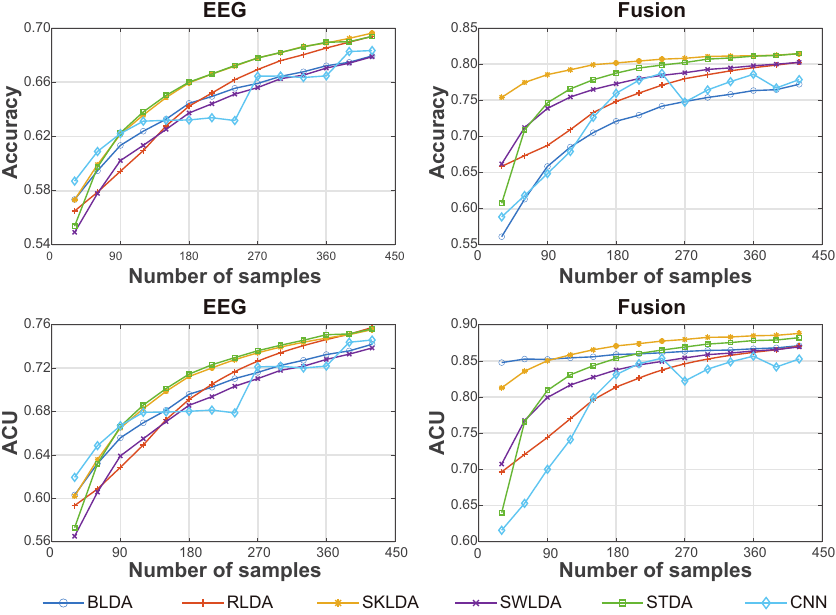}
\caption{Comparison of Accuracy and AUC across the employed methods under different number of training samples.}
\end{figure}
\begin{figure}[htbp]{\label5}
\centering
\includegraphics[scale = 1]{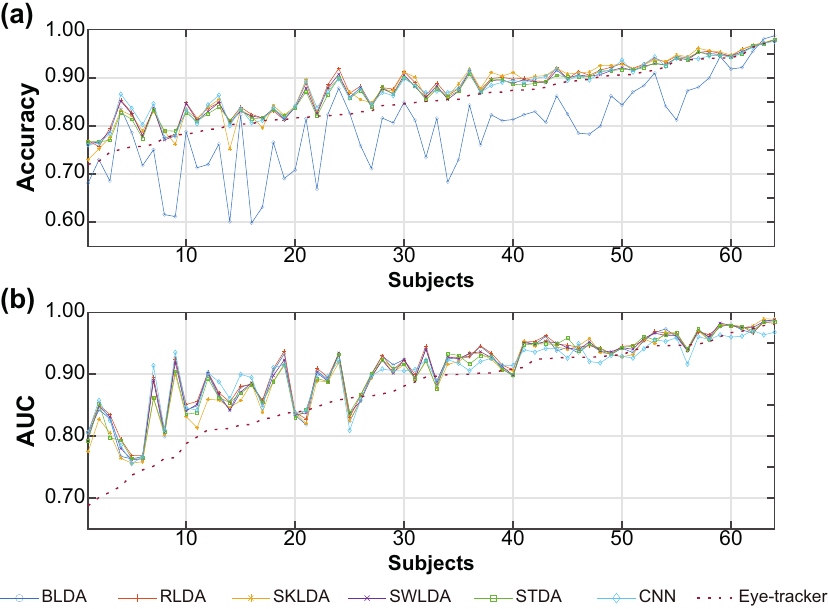}
\caption{(a) Accuracy and (b) AUC for multiple classifiers of each subject. The subject was sorted by the ascending of classification performance based upon the eye-tracker model.}
\end{figure}

\begin{table*}[hbtp]
\centering
\begin{threeparttable}
\caption{Offline Classification Performance across Different Algorithms}
\begin{tabular}{lllll}
\hline \hline
\multirow{2}{*}{Algorithms} & \multicolumn{2}{c}{EEG} & \multicolumn{2}{c}{Fusion}\\
\cline{2-5}
 &  Accuracy & AUC & Accuracy & AUC\\
\hline
RLDA & 0.7793 $\pm$ 0.0403 & 0.8007 $\pm$ 0.0599 & 0.8802 $\pm$ 0.0551 & 0.9104 $\pm$ 0.0547\\
SWLDA & 0.7754 $\pm$ 0.0382 & 0.7908 $\pm$ 0.0617 & 0.8761 $\pm$ 0.0534 & 0.9072 $\pm$ 0.0565\\
BLDA & 0.6716 $\pm$ 0.0504 & 0.7933 $\pm$ 0.0632 & 0.7954 $\pm$ 0.0910 & 0.9066 $\pm$ 0.0573\\
SKLDA & 0.7225 $\pm$ 0.0503 & 0.7838 $\pm$ 0.0615 & 0.8783 $\pm$ 0.0603 & 0.9004 $\pm$ 0.0608\\
STDA & 0.7652 $\pm$ 0.0367 & 0.7789 $\pm$ 0.0583 & 0.8740 $\pm$ 0.0544 & 0.9049 $\pm$ 0.0568\\
CNN & 0.7854 $\pm$ 0.0403 & 0.8053 $\pm$ 0.0622 & 0.8772 $\pm$ 0.0535 & 0.9028 $\pm$ 0.0512\\
\hline \hline
\end{tabular}
Note: Values are presented as mean $\pm$ S.D., Fusion indicates features from EEG and eye-tracker were fused to obtain the Accuracy and AUC.
\end{threeparttable}
\end{table*}

\subsection{Offline Classification}
In the offline classification, we first assessed the performance of classifiers under different number of training samples. In line with previous study~\cite{xiao2019discriminative}, we found that the classification performance was monotonically increased with the number of training samples ({\color{blue}{Fig. 4}}). We then assessed the classification performance when using features from eye-track and EEG data respectively. 
When using eye-track feature, we obtained the classification accuracy of 0.8527 $\pm$ 0.0638 and the AUC of 0.8734 $\pm$ 0.0746 that was served as benchmark. However, the classification performance across different algorithms using only EEG data is significantly lower than the benchmark probably due to the large inter-individual differences in single-trial EEG characteristics ({\color{blue}{Table I}}). Moreover, we found that through employing the features from both eye-track and EEG data, the classification performance was significantly improved and exhibited a superior performance compared to the benchmark for most of the subjects ({\color{blue}{Fig. 5}}). Further interrogation of the classification performance across six methods, we found BLDA exhibited relatively low accuracy in the fusion manner. Hence, the remaining five methods (i.e., RLDA, SWLDA, SKLDA, STDA, and CNN) were selected for the following pseudo-online validation.   

\begin{figure}[htbp]{\label6}
\centering
\includegraphics[scale = 1]{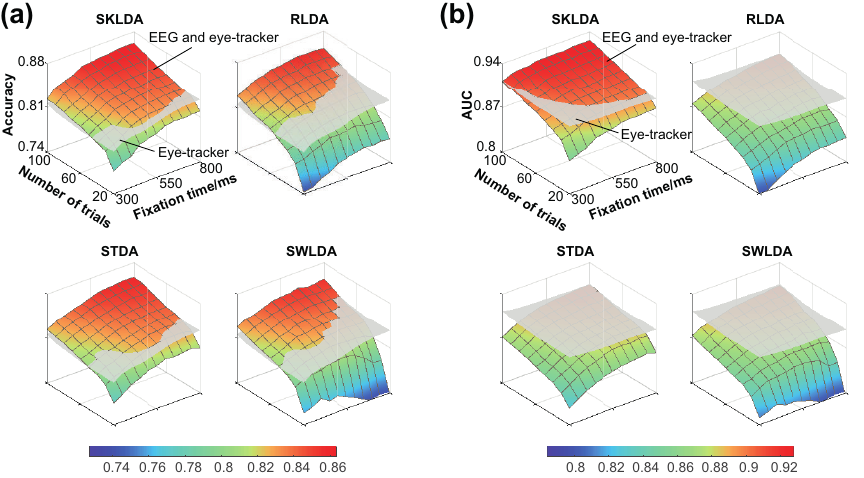}
\caption{Distribution of (a) Accuracy and (b) AUC across employed methods that exhibited different performance under number of training trials and durations of fixation. The gray surface indicated the performance when only using eye-tracking data that consider as the benchmark.}
\end{figure}

\subsection{Pseudo-Online Classification}
The performance of the pseudo-online classification was shown in ({\color{blue}{Table II}}). Again, we first obtained the performance benchmark from eye-track data (accuracy = 0.8277 $\pm$ 0.0862 and AUC = 0.9035 $\pm$ 0.0760). Similar to the offline results, the classification performance with single EEG feature was significantly lower than the benchmark for all methods. Although the performance was greatly improved by adding eye-tracker feature into classification, only SKLDA obtained both higher accuracy (0.8454 $\pm$ 0.0868) and higher AUC (0.9171 $\pm$ 0.0614) in a single-trial classification time of 2.3343 $\pm$ 0.0306 ms. Of note, the performance of CNN was significantly lower than the other four methods with a longer classification time. Following validation was therefore only performed on the remaining four methods.  

\begin{table*}[htbp]
\centering
\begin{threeparttable}
\caption{Classification Performance across Different Algorithms during Online Validation}
\begin{tabular}{lllllll}
\hline \hline
\multirow{2}{*}{Algorithms} & \multicolumn{3}{c}{EEG} & \multicolumn{3}{c}{Fusion}\\
\cline{2-7}
 &  Accuracy & AUC & Time (ms) & Accuracy & AUC & Time (ms)\\
\hline
RLDA & \small{0.6550 $\pm$ 0.0560} & \small{0.7125 $\pm$ 0.0692} & \small{3.3722 $\pm$ 0.0544} & \small{0.8419 $\pm$ 0.0879} & \small{0.8890 $\pm$ 0.0778} & \small{3.2570 $\pm$ 0.0464}\\
SWLDA & \small{0.6466 $\pm$ 0.0535} & \small{0.7061 $\pm$ 0.0669} & \small{2.3832 $\pm$ 0.0243} & \small{0.8388 $\pm$ 0.0910} & \small{0.8858 $\pm$ 0.0820} & \small{2.3198 $\pm$ 0.0326}\\
SKLDA & \small{0.6595 $\pm$ 0.0558} & \small{0.7624 $\pm$ 0.0513} & \small{2.3773 $\pm$ 0.0348} & \small{0.8454 $\pm$ 0.0868} & \small{0.9171 $\pm$ 0.0614} & \small{2.3343 $\pm$ 0.0306}\\
STDA & \small{0.6543 $\pm$ 0.0568} & \small{0.7123 $\pm$ 0.0674} & \small{3.2896 $\pm$ 0.0687} & \small{0.8410 $\pm$ 0.0855} & \small{0.8914 $\pm$ 0.0791} & \small{3.2217 $\pm$ 0.0287}\\
CNN & \small{0.6592 $\pm$ 0.0557} & \small{0.7215 $\pm$ 0.0661} & \small{5.2227 $\pm$ 0.1007} & \small{0.7293 $\pm$ 0.0941} & \small{0.8426 $\pm$ 0.0790} & \small{6.2242 $\pm$ 0.0983}\\
\hline \hline
\end{tabular}
Note: Time indicates the duration for single-trial classification. 
\end{threeparttable}
\end{table*}

To investigate how the number of training trials and the duration of fixation influence the pseudo-online classification performance, the distribution of accuracy and AUC with different settings for these four selected classifiers was shown in ({\color{blue}{Fig. 6}}). The rendered surface represented the classification performance obtained with fusion features, while the gray surface indicated the performance when only using eye-tracking data that consider as the benchmark. Similar to the offline results, the performance of all methods improved monotonically with increasing number of training trials. In contrast, the classification performance exhibited a complex dependent level for the setting of fixation duration, i.e., the best performance was not always obtained using long fixation duration. Among the four methods, only SKLDA exhibited superior classification performance in most of the settings compared to the benchmark. 

\section{Discussion}
In this study, we revealed an explicitly improved performance with the fusion of ERP and eye-tracking data in the single trial classification of free search task, both in offline and online analysis. Our previous research proved the effectiveness of block highlight display (BHD) eye-controlled technique~\cite{ge2021using}, which was further validated by the high target-detection rate in the performance of active search paradigm in this study. The ERP components were also demonstrated in different spatial (occipital area) and temporal (100 and 300 ms) characteristics between target and non-target detection. Six widely-used classifiers were employed to verify the effectiveness of the hybrid BCI system. In the offline analysis, the classification approach with multi-modal inputs of fixation duration and ERP significantly outperformed the method with single-modal brain/eye features. Besides, the multiple versions of LDA approach, except for BLDA, were proved effective in the single-trial classification, showing a superiority in solving such problem with relatively simple inputs than sophisticated neural network structure. As for the online validation, we found the fused feature still provided a robust performance, while different classification approaches exhibited divergent adaptability towards limited number of samples and response time. And SKLDA ranked the top among all LDA classifiers from both the accuracy and application efficiency.

In the offline analysis, we observed that the fused feature provided a much higher accuracy and AUC in the 10-fold classification. When ERP was taken as the only feature, the classification performance of different classifiers was scattered. Although ERP signal could reflect cognitive processes with a high temporal resolution, its low amplitude and sensitivity to various artifacts made it hard to be extracted stably and also varied across different subjects. This was consistent with previous findings that the prediction performance using ERP was mixed in different studies. Specifically, SKLDA was proposed for single-trial classification of ERP-based BCI by Blankertz et al.~\cite{blankertz2011single}, suggesting superior performance over ordinary LDA and SWLDA. Zhang developed STDA and demonstrated its superiority among several forms of LDA methods (LDA, SWLDA and SKLDA) in ERP classification~\cite{zhang2013spatial}. In our study, we found that among the LDA-based classifiers, RLDA outperformed other methods with an accuracy of 0.7793 and BLDA ranked the last with an accuracy of 0.6716. As the most popular deep learning method in ERP-related studies~\cite{craik2019deep}, CNN also has an exceptional performance with the top accuracy of 0.7854. Briefly, the performance of classifier varied across different studies and datasets for ERP-based classification.

However, when ERP and eye-tracker data were collectively adopted for classification input in the present study, the performance of divergent classifiers was greatly improved than only taking single-model ERP or eye-track feature. Besides, it was interesting to find that the accuracy and AUC of different classification algorithms became similar. In our experiment protocol, fixation duration tended to be higher when participants gazed on the target icon, consonant with the practical scenarios. The eye-tracker ensures a high temporal and spatial accuracy towards the gazing time and position, indicating a robust measure of underlying cognitive processes based on eye movement-related variables, such as fixation duration and saccade~\cite{baccino2011eye}. Therefore, compared with the single ERP signal, the fuse of fixation duration from eye-tracker provided a relatively stable criterion towards cognitive states without large inter-subject variability, so that the input formulation was highly adaptable and the prediction performance tended to be similar across different algorithms. On the other hand, when compared with the single fixation duration, the accuracy was also improved by feature fusing, which demonstrated that ERP was associated with ongoing intention of target selection to enhance the recognition performance of traditional eye-tracking system. Among different classification approaches, the performance of RLDA, SKLDA, SWLDA and STDA were close with the accuracy of 0.8802, 0.8783, 0.8761 and 0.8740 respectively. In addition, in the 10-fold offline analysis, CNN ranked the second when fusing eye and brain features, which was possibly due to the abundant training sample and unlimited processing time. When training set was massively cut down, most LDA-based classifiers outperformed CNN with fused feature. Furthermore, in the profile of classification performance with increasing training sample, we found that the SKLDA, STDA, SWLDA and RLDA provided an improving and robust accuracy for the fusion approach, and SKLDA was extraordinarily outstanding over other classifiers under various scale of training set.

According to the review of Lotte et al, LDA is one of the most prevalent forms of classifiers for EEG-based BCI, especially for online and real-time processing~\cite{lotte2018review}. The online analysis in the present study further validated the improvement effect of feature fusion in the selected classification algorithms. Notably, we tested the practicability of the hybrid system by modulating the amount of training set and the length of every input sample (i.e. system response time). Firstly, consistent with offline analysis, the accuracy and AUC maintained at a high level when 80 trials were taken for training and decreased with the contraction of training set. The amount of training sample determined the initial calibration time of the system and were directly related to the interaction convenience, but an extremely small training set, like 20 trials, also deteriorated the classification performance immensely. In addition, in order to investigate the effect of system response time to the decoding of interaction intention, we truncated the length of fixation duration and corresponding ERP epoch of each sample, for the sake of simulating different response time for intention recognition of the hybrid system and evaluating the classification performance. The accuracy and AUC declined when response duration decreased, with a slight slope from 800 ms to 400 ms and a substantial drop below 400 ms. From the perspective of eye-tracking data, using a shorter recognition time was harder to distinguish the intended selection from other conditions, such as long processing time towards stimulus for participants. Previous study showed that the dwell time of novices was typically between 450 and 1000 ms in gaze typing tasks, and decreased to 282 ms after repeated training~\cite{majaranta2009fast}. In other eye selection related ERP studies, the fixation duration were usually set in a long threshold, such as 1000ms~\cite{protzak2013passive} or 2000ms~\cite{zander2011towards}. Considering no pre-training was performed on our subjects, a proper recognition threshold above 400 ms could ensure a more accurate performance. From the aspect of ERP, it was obvious to find out that shorter EEG epoch after stimulus comprising less ERP components. Given that the predominate ERP was concentrated on the 100ms and 300 ms as shown in ({\color{blue}{Fig. 3}}), the epoch less than 400ms might induce the loss of crucial information especially when latency shift happened in P300 wave. However, the superiority of the fused feature was still observed compared with single fixation duration across the whole range of system response time (300 to 800 ms) as long as considerable training set was implemented. Single eye-tracker was natural to use and could reach a decent accuracy without much training, but only by a few calibrations, the fusion of ERP and fixation duration could outperform the former. Besides, consistent with offline analysis, SKLDA still maintained the best performance for the fused feature among all classification methods, showing the largest area over single eye-tracker classification performance in ({\color{blue}{Fig. 5}}), and a relatively less requirement of training set and response duration for the outperformance. Previous study proved that SKLDA remained effective when training set was insufficient~\cite{blankertz2011single}, so that it was practical to be utilized in real-time scenarios. As for the computational time, we found that the single-trial process time was associated with the length of input signal, especially in RLDA. And the classifiers of SKLDA and SWLDA required less computation cost in the practical applications.

In the present study, some factors should be considered when interpreting our results. Firstly, only young and healthy university students were included as the sample to test our protocol. But there is evidence suggesting that the performance of LDA declined largely for the ERP classification of the severely disabled in real-life applications~\cite{martinez2016asynchronous}. Further study could extend the diversity of subjects to verify the robustness of fused features. Secondly, in the design of present interaction interface, the icons were arranged isolated and regularly with rigid distance as an array, to which the users would be accustomed with the progress of experiment. A real ecological application was supposed to display arbitrary targets, which provided a more oddball stimuli to the user and elicited stable ERP responses. At the same time, the stimulus was expressed in images in this study, while other type of interaction item, like words, was associated with different predominant ERP components~\cite{baccino2011eye}. Future study could introduce different forms of selection target to test the generalization of the hybrid system. In addition, this study observed a relatively limited increase of performance by fusing eye and ERP compared with single eye-tracker signal as the feature. Considering the computational complexity, only the time domain wave was taken from ERP in our pipeline. Other features could also be explored to overcome the shortcomings in ERP characteristics and optimize the performance of the hybrid interaction system for single trial classification.

\section{Conclusion}
In the current study, we introduced an eye-brain hybrid BCI interaction system and assessed the performance in a customized free visual search paradigm. In comparison with the single-model EEG or eye-track features, the proposed hybrid BCI system achieved better performance in both offline and online conditions. Furthermore, practical validation across six widely used classification methods showed that the SKLDA method could maintain superior performance under condition with few training set and fast response time. In sum, our study shed new insights on the approach of hands-free HCI and provided novel and practical solution to the intention detection in the real-world scenarios.



%



\ifCLASSOPTIONcaptionsoff
  \newpage
\fi

\bibliographystyle{IEEEtran}
\bibliography{references}

\end{document}